\documentclass[aps,prl,twocolumn,superscriptaddress,amsmath,amssymb]{revtex4}
\usepackage{graphics,graphicx}

\begin{document}

\title{Pressure-induced Spin-Peierls to Incommensurate Charge-Density-Wave Transition in the Ground State of TiOCl}

\author{A.~Prodi}
\affiliation{Department of Physics, Massachusetts Institute of
Technology, Cambridge, MA 02139, USA}

\affiliation{Center for Materials Science and Engineering,
Massachusetts Institute of Technology, Cambridge, MA 02139, USA}
\author{J.S.~Helton}
\affiliation{Department of Physics, Massachusetts Institute of
Technology, Cambridge, MA 02139, USA}
\author{Yejun~Feng}
\affiliation{The Advanced Photon Source, Argonne National Laboratory, Argonne, IL 60439, USA}
\author{Y.S.~Lee}
\affiliation{Department of Physics, Massachusetts Institute of
Technology, Cambridge, MA 02139, USA}

\date{\today}

\begin{abstract}
The ground state of the spin-Peierls system TiOCl was probed using
synchrotron x-ray diffraction on a single-crystal sample at $T=6$~K.
We tracked the evolution of the structural superlattice peaks
associated with the dimerized ground state as a function of
pressure.  The dimerization along the $b$ axis is rapidly suppressed
in the vicinity of a first-order structural phase transition at $P_C
= 13.1(1)$~GPa.  The high-pressure phase is characterized by an
incommensurate charge density wave perpendicular to the original
spin chain direction.  These results show that the electronic ground
state undergoes a fundamental change in symmetry, indicating a
significant change in the principal interactions.
\end{abstract}


\maketitle

\begin{figure}[top]
\includegraphics[width=3.3in]{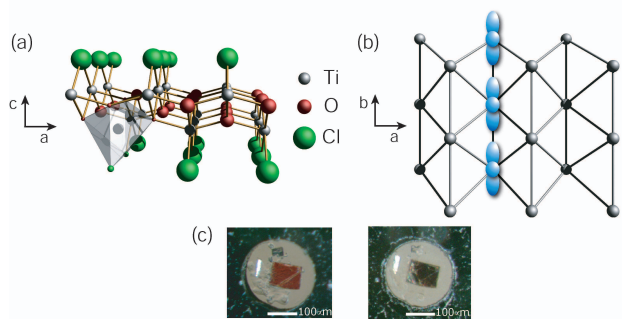}
\caption{\label{fig:struct}(Color Online) (a) Crystal structure of TiOCl viewed along
the spin chain direction.  The Ti atoms are shown in gray. (b) Ti
sublattice showing the orbital arrangement yielding chains along $b$
in the low-pressure phase. (c) Micrograph of the TiOCl sample inside
the diamond anvil cell at room temperature, before (with $P$$<$0.5~
GPa) and after (with $P\sim3$~GPa) the x-ray measurements.
The silver pressure calibrant is visible at the left of the sample.}
\end{figure}

Low-dimensional quantum magnets display rich physics due to the
interplay of spin, lattice, orbital, and charge degrees of freedom.
As a well-known example, the spin-Peierls transition occurs when the
presence of the spin-phonon coupling in spin 1/2 antiferromagnetic
chains causes the formation of singlet pairs on structural dimers.
Only a handful of physical systems have been identified with a
spin-Peierls ground state, including several complex organic
compounds \cite{PhysRevB.14.3036} and the inorganic compound
CuGeO$_3$ \cite{PhysRevLett.70.3651}.  However, the latter material
may not be a good realization of a conventional spin-Peierls system
\cite{PhysRevLett.70.3651,PhysRevB.51.16098,PhysRevB.53.5579}.
Recently, the inorganic material TiOCl has been shown to exhibit
spin-Peierls
physics \cite{PhysRevB.67.020405,PhysRevB.68.140405,hoinkis:245124,shaz:100405,abel:214304}.
The system is composed of chains of Ti ions which undergo successive
transitions upon cooling to an incommensurate nearly dimerized state
at $T_{C2}=92$~K and to a commensurate dimerized state at
$T_{C1}=66$~K, respectively. The orthorhombic crystal structure of
TiOCl, illustrated in Fig.~1, consists of bilayers of Ti and O
atoms, separated by chlorine layers along the $c$-axis. Within the
$a-b$ plane, Ti$^{3+}$ (3$d^1$, $S$=$ \frac{1}{2}$) ions form a
buckled rectangular lattice where the dominant magnetic interaction
is due to Ti-Ti direct exchange along the $b$-direction. This gives
rise to one-dimensional quantum spin chains along $b$, where the
interchain interaction is frustrated. The geometry of the Ti ions is
depicted in Fig.~\ref{fig:struct}(b). The formation of a dimer state
below $T_{C1}=66$~K involves a substantial displacement ($
\delta$=0.03 $b$) of Ti$^{3+}$ ions along the chain direction,
doubling the unit cell along the $b$-axis with the appearance of
superlattice reflections with wavevector
$\mathbf{q}=(0,\frac{1}{2},0)$\cite{shaz:100405,abel:214304}. A key
prediction of the conventional spin-Peierls theory
\cite{PhysRevB.19.402} has been confirmed in TiOCl with the direct
observation of the softening of the spin-Peierls active phonon at
$T_{C1}$ by means of inelastic x-ray scattering \cite{abel:214304}.

Hydrostatic pressure has served as a key experimental tuning
parameter in exploring the interconnection between different
correlated electronic states. Rich phenomena driven by pressure in
low-dimensional systems include commensurate-to-incommensurate
transition in charge-density-wave dichalcogenides \cite
{PhysRevLett.45.269,PhysRevB.23.2413,PhysRevB.19.1610}, spin-Peierls
to superconductivity transition in organic systems
\cite{organic:08}, a hierarchical phase diagram in the spin-ladder
compound NaV$_2 $O$_5$\cite{PhysRevLett.87.086402,ohwada:094113},
and novel phase transitions in CuGeO$_3$ \cite
{PhysRevLett.77.1079,PhysRevLett.78.487}.  Recent studies of TiOCl
under pressure \cite{kuntscher:035106} have suggested that the
system undergoes a metallization associated with changes in the
crystal structure at pressures $P\sim16$~GPa. The intepretation of
these results in terms of bandwidth-controlled Mott transition is
still under debate \cite {forthaus:165121,zhang:136406}. Measurements
of the electrical resistance under pressure show that TiOCl remains
semiconducting up to 24 GPa, however, the energy gap indicates a
critical pressure of $\sim13$~GPa above which the electronic
structure changes \cite {forthaus:165121}. Currently, there are no
measurements of how the dimerized spin-Peierls ground state is
affected by increasing pressure through the critical pressure. Most
previous experiments were carried out at room temperature and are
not sensitive to the low-temperature ground state.

We report direct measurements of both the fundamental lattice and
weaker superlattice structural reflections in TiOCl under pressure
and at low temperatures using high-resolution x-ray diffraction on a
single crystal sample.
We observed a sudden first-order structural phase transition and
determined the transition pressure to be $P_C = 13.1$~GPa.  In       
addition, the superlattice peaks indicate that the low pressure
spin-Peierls dimerization is significantly diminished at the phase
boundary and disappears above $P_C$.  In the high pressure phase,
the system orders with an incommensurate charge density wave along
the $a$-axis direction (perpendicular to the spin-chains of the low
pressure phase). This new feature of the ground state points to
either a change in the dimensionality of the high pressure state or
a significant change in the principal interactions.

Synchrotron x-ray diffraction experiments at high-pressures and low
temperatures were carried out at the 4-ID-D beamline of the Advanced
Photon Source at Argonne National Laboratory.  Incident x-rays with
energy 20 keV were selected using a Si(111) double monochromator and
were focused onto the sample by Pd mirrors. A He-gas driven membrane
diamond anvil cell was mounted on the cold finger of a
closed-cycle cryostat to apply pressures up to 16 GPa with
\textit{in situ} tunability \cite{feng:137201,jaramillo:184418}.
Measurements were performed by monotonically increasing the membrane
pressure in small ($\sim0.2$~GPa) increments while at base temperature
$T_0=6.0 \pm 0.5$~K.  The cryostat was installed on the sample stage
of a psi-circle diffractometer working in the vertical scattering geometry.
The high collimation of both the incident and scattering x-ray paths, the
large ($>$1m) Rowland circle, and the use of a NaI point detector,
yield high wave vector resolution and efficient rejection of background
scattering from the sample environment. Pressure was calibrated
\textit{in situ} \cite{jaramillo:184418} against the lattice
constant of silver, determined from powder lines of a silver foil
inside the pressure chamber.

\begin{figure}[top]
\includegraphics[width=2.5in]{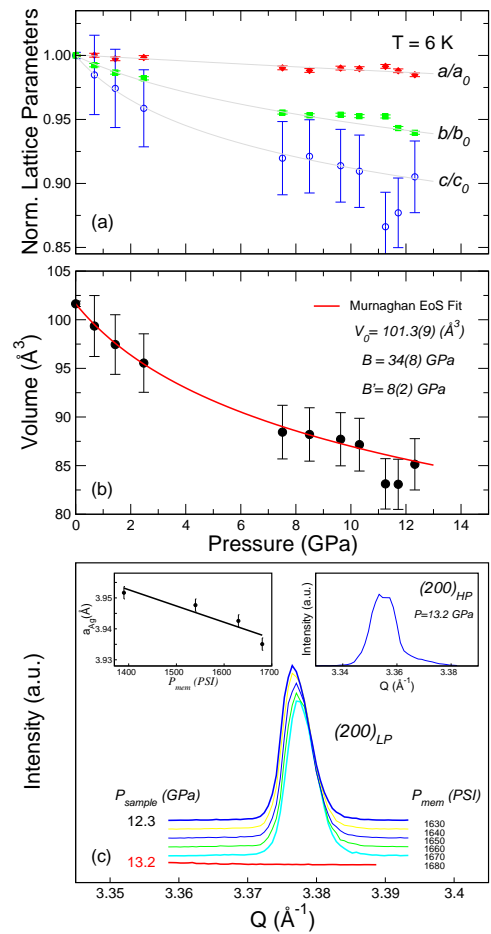}
\caption{(Color Online) (a) Pressure dependence of the unit cell
parameters of TiOCl at $T=6$~K. (b) Low temperature unit cell volume
with fit to a Murnaghan equation of state. (c) Measurements of the
(200) fundamental Bragg reflection as the pressure is increased in
fine steps.  Left inset: the sample pressure measured by the Ag
lattice constant is linear relative to the membrane load in this
pressure range.  Right inset: the (200) peak of the high-pressure
phase, using pseudo-orthorhombic notation.}
\end{figure}

Single crystals of TiOCl of 100 $\mu$m $\times$ 100 $\mu$m $\times$
10$\mu$m typical size and 0.3 $^{\circ}$ mosaic spread were selected
from batches grown by the vapor transport method
\cite{PhysRevB.67.020405}. A 4:1 methanol/ethanol mixture was used
as the pressure transmitting medium.  The flaky, amber colored
single crystal lay flat on the diamond culet when mounted, with the
$c$-axis oriented along the loading axis. In the resulting
(transmission) scattering geometry, diffraction peaks are
easily accessible in the $L = 0$~plane, while information regarding the
$c$-axis lattice constant can be accessed via ($H, 0, L$)
reflections.
In total, three fundamental Bragg ((200), (020), and (201)) 
and five superlattice reflections were observed as a function of pressure.
We note that during the experiment, the width of one of the
rocking scans increased under pressure from 
0.3$^{\circ}$ at 0.8 Gpa to $\sim5^{\circ}$ at 12 GPa.
This increased mosaic spread likely
results from a slight crumpling of the thin crystal under pressure in the
diamond anvil cell.  The crystal, however, remains intact, and the
positions of the Bragg peaks and the relative intensities of the
superlattice peaks can still provide important information, as we
discuss below.  We also note that the alcohol mixture has been
proven to provide static pressure conditions with homogeneity better
than $\Delta$P= 0.1 GPa in this pressure and temperature range
\cite{jaramillo:184418}.

\begin{figure}[top]
\includegraphics[width=3.0in]{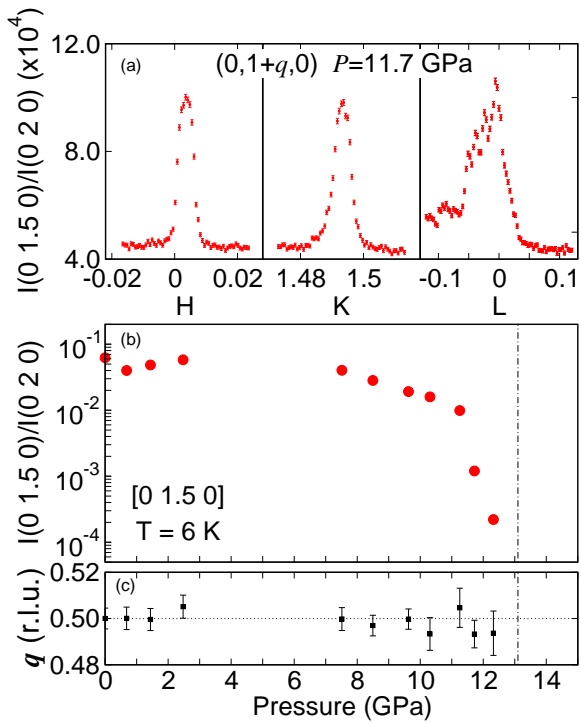}
\caption{\label{fig:slvsp_tmp} (Color Online) Pressure dependence of
the (0 $\frac{3}{2}$ 0) superlattice reflection of TiOCl at $T=6$~K.
(a) Representative scans at 11.7 GPa; the intensity is scaled to the
closest fundamental Bragg peak (0 2 0). (b) Intensity and (c) wave
vector \textit{\textbf{q}} of the structural superlattice as a
function of pressure. The error bars for the intensity is smaller
than symbol size, while the error bars for \textit{\textbf{q}}
represent the fitted widths of the scans along $K$.}
\end{figure}

We first display the measured TiOCl lattice parameters along all
three axes as a function of pressure in the low temperature
spin-Peierls phase at $T=6$~K as shown in Fig.~2. The lattice
parameters are normalized to their corresponding values measured at
ambient pressure.
The zero-pressure axial compressibilities,
$\beta_{0,i}$=$a_{0,i}^{-1}(\partial a_i/\partial P)_{P=0}$,
show a marked anisotropy with
$\beta_{0,a}$:$\beta_{0,b}$:$\beta_{0,c}\approx$ 1:5.3:10,
$\beta_{0,a}$=$1.245(1)\times 10^{-3}$ GPa$^{-1}$.
A least squares fit of the unit cell volume to a Murnaghan
Equation of State \cite{Angel:01} leads to a bulk modulus of $B_0$=34
GPa and $B'_0$=8 GPa.  These values are consistent with the
corresponding values reported from powder x-ray diffraction at room
temperatures \cite{kuntscher:241101,kuntscher:035106}.  At $P$=12.3
GPa we find $a$=3.729(5) \AA, $b$=3.141(4) \AA\ and $c $=7.27(22)
\AA\ as referred to the pseudo-orthorhombic setting.

As shown in Fig.~2(c), a first-order structural transition is
observed between low pressure (LP) and high pressure (HP) phases,
indicated by the sudden disappearance of the (200)$_{LP}$ reflection.
A precise value of $P_c$=13.1(1) GPa  for the transition pressure can
be obtained by finely stepping the membrane pressure and monitoring the
(200) fundamental Bragg reflection.The left inset shows that the measured
lattice constant of the Ag foil varies linearly with pressure in this range.
Above $P_c$, three new peaks belonging to the high-pressure phase could be
indexed as (200)$_{HP}$, (020)$_{HP}$ and (201)$_{HP}$ again in a
pseudo-orthorhombic unit cell setting with $a$=3.746(11) \AA,
$b$=3.150(8) \AA\ and $c$=6.94(57) \AA.  It is interesting to note
that the $a$ lattice parameter \textit{increases} from the
low-pressure to high-pressure structures, a result predicted by
recent \textit{ab initio} calculations \cite{zhang:136406}.  The
volume strain associated with the transition is $V_s$ = ($V_{HP}$ -
$V_{LP}$)/$V_{LP}$ = -0.0374.  After warming the cell to room
temperature, visual inspection evidenced a color change in the
sample, shown in Fig.~1(c). The dark brown color of the high
pressure sample is retained upon decompression down to 3 GPa (the
color is similar to that observed in the ``metallic'' state at room
temperature reported previously \cite{kuntscher:241101}).  This
large pressure hysteresis is striking, though some degree of
hysteresis is expected for a first-order structural transition.

\begin{figure}[top]
\includegraphics[width=3.0in]{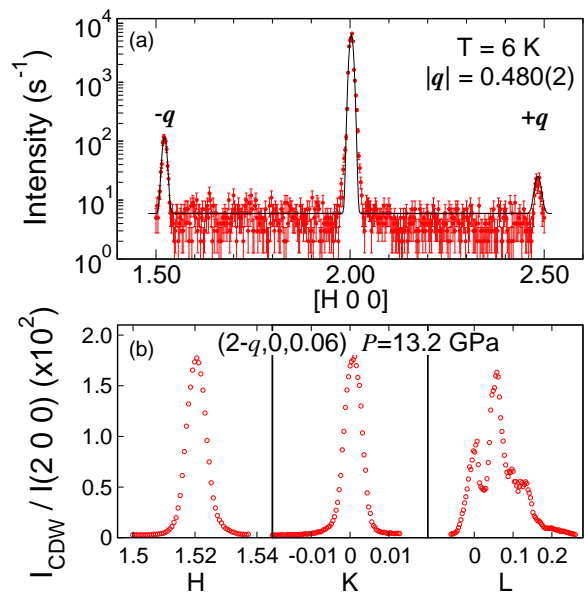}
\caption{\label{fig:HP-IC} (Color Online) (a) Longitudinal scan
along the [$H00$] direction of the high pressure phase of TiOCl at
$T=6$~K.  A new pair of superlattice peaks are observed around the
(200) Bragg peak with incommensurate wave vector $q=0.48$. (b) Scans
through the incommensurate (2-$q$, 0, 0) peak along the $H$, $K$, and
$L$ directions, using pseudo-orthorhombic notation.}
\end{figure}

One of the main results of this work is the observation of the weak
diffracted signal from the superlattice peaks at pressures above and
below $P_C$.  For pressures below $P_C$, we measured the (0,1.5,0)
superlattice peak which reflects the dimerized structure of the
spin-Peierls phase.  The data are shown in Fig.~3.  In panel (a), we
show $H$, $K$, and $L$ scans through the peak, where the intensity
is normalized to that of the nearby (020) Bragg peak.  Note that
that this superlattice peak (which is one of the most intense) is
about three orders of magnitude less intense than the fundamental
Bragg peak, hence it is difficult to observe in measurements on
powder samples. 
The pressure dependence of the normalized intensity of the (0
$\frac{3}{2}$ 0) superlattice reflection is shown in Fig.~3(b).  The
intensity remains relatively flat over a wide pressure range up to
11 GPa, then rapidly decreases over a pressure range of about 2 GPa
prior to reaching $P_C$.  The modulation wavevector
\textit{\textbf{q}}, shown in panel (c), indicates that the periodic
distortion remains commensurate with the lattice at all pressures
(within alignment errors).  We also find that the superlattice peak
remains resolution-limited, within the errors, at all pressures
along $H$ and $K$, indicating long-range order of the dimerized
phase exists up to $P_C$. The other measured superlattice
reflections (not shown) display similar behavior. Since the
diffracted intensity is related to the modulation as $I \propto
|\roarrow Q \cdot \roarrow \delta|^2$ and the largest contribution
is due to the Ti atomic displacement along the $b$-axis, we can
estimate that the Ti-Ti dimerization amplitude $\delta$ is
suppressed by at least a factor of 15 prior to reaching the first
order phase transition.

Interestingly, in the high pressure phase, a new ground state
emerges which is characterized by a pair of incommensurate
superlattice peaks at (1.52, 0, 0) and (2.48, 0,0) as shown in Fig.
4.  These peaks are near the commensurate positions ($2 \pm
0.5,0,0$), and the incommensurate nature of this new structure
can be clearly seen in the longitudinal scan of Fig.~4(a).  We can describe
this modulation of the high-pressure structure with a modulation
vector \textit{\textbf{q}} $=$ (1 -
$\varepsilon$)\textit{\textbf{a}}$^{\ast}$/2 and discommensuration
$\varepsilon$ = 0.02.  Surprisingly, instead of ordering along the
$b$-axis as in the spin-Peierls ground state, the new modulation
vector is along the $a$-axis, which is perpendicular to the
low-pressure spin chain direction.  This is in contrast to the
incommensurate structure of TiOCl observed at ambient pressure
for temperatures between $T=66$~K to 93~K
\cite{abel:214304,krimmel:172413}, which centers around the ($0, 2
\pm 0.5,0$) position.  We did not observe any superlattice
diffraction intensity at ($0, 2 \pm 0.5,0$) in the high-pressure
phase.

The stability of the spin-Peierls ground state results from the
balance between gain in magnetic energy from the dimerization and
the cost in elastic energy of the associated distortion.  The
competing terms are tuned simultaneously by pressure: while the
reduction of interatomic distances upon compression always increases
the elastic energy cost, the influence on the magnetic energy is
more subtle through its dependence on the exchange $J$, the
spin-phonon coupling constant, and on the soft phonon mode.  Our
results suggest that the balance holds until a threshold is reached
near $P_C$ and the dimerization is rapidly suppressed.  A recent
DMRG study suggests that the dimerization amplitude $\delta$ is
independent of $J_{a,c}/J_b$ for a wide range of the
parameters \cite{0953-8984-20-13-135223}. In the high pressure phase,
the incommensurate structure along the $a$-axis may originate from a
Fermi-surface nesting of an itinerant electron state. A previous
report using powder x-ray diffraction study claimed that a
conventional Peierls dimerization exists along the $b$-direction in
the high-pressure phase at room temperature \cite{blancocanosa-2008}.
Our results show that this proposed state does not exist at $T=6$~K.
If conventional Peierls physics is operative at high-pressure, then
the chain direction should have switched to the $a$-direction.
The slight incommensurability that we observe would be hard to
explain in such a model. One possibility is that the dimensionality
of the system may change from one-dimensional to quasi
two-dimensional when crossing from the insulating low-pressure phase
to the more metallic high-pressure phase. Further work is needed to
confirm the charge-density wave nature of the ground state, such as
searching for thermodynamic signatures or Kohn anomalies.

In summary, by performing high pressure x-ray scattering studies of
a single crystal sample of TiOCl at low temperatures, we directly
observed the diminishment of the dimerized superlattice in the
vicinity of the critical pressure $P_C$.  The spin-Peierls
dimerization along $b$ does not survive above $P_C$. Above $P_C$,
incommensurate superlattice peaks along the $a$ direction are
observed.  We conclude that a charge density wave exists in the
ground state of the high-pressure phase.

\begin{acknowledgments}
We thank F.C.~Chou, S.H.~Shim, E.T.~Abel and D.B.~McWhan for
fruitful discussions.  The work at MIT was supported by the
Department of Energy (DOE) under Grant No.~DE-FG02-07ER46134.  Use
of the Advanced Photon Source at Argonne National Laboratory was
supported by the DOE under Contract No.~DE-AC02-06CH11357. This work
used facilities supported in part by the NSF under Agreement No.
DMR-0454672.
\end{acknowledgments}

\end{document}